\documentclass{taiwan}
\def\etal{{\it et al.}}
\begin{document}

\title{Future $B$ Experiments from The BTeV/LHC-b Perspective}

\author{SHELDON STONE }

\address{Physics Department, Syracuse University, Syracuse N. Y., 
USA, 13244-1130\\
E-mail: stone@phy.syr.edu}  


\maketitle

\abstracts{Many measurements are necessary in the program of studying mixing, 
CP violation and rare decays of $b$ and $c$ quarks. These measurements require
large numbers of $B^o$, $B_s$, $B^-$ and $D^{*+}$ hadrons. Fortunately,  
copius production of particles containing $b$ and $c$ quarks will occur at
Tevatron and the LHC. The crucial measurements are described here, as well as 
the design of the two experiments, LHC-b and BTeV, that can exploit the
$4-20~\times 10^{11}$ $b$ hadrons produced every $10^7$ seconds.}

\section{Introduction}

The basic experimental goals are to make exhaustive search for physics beyond the Standard Model
and to precisely measure Standard Model parameters. 
I first discuss what studies need to be done, not just what studies can be done
in the near future. 
Measurements are necessary on CP violation in $B^o$ and $B_s$ mesons, $B_s$
mixing, rare $b$ decay rates, and mixing, CP violation and rare decays in the
charm sector.

\section{The CKM Matrix and CP Violation}
\label{sec:Intro}
\subsection{The 6 Unitarity Triangles}
The base states of quarks, the mass eigenstates, are mixed to form the
weak eigenstates as described by the $3\times 3$ complex
Cabibbo-Kobayashi-Maskawa matrix,\cite{ckm} that can be expressed in terms of 4 
fundamental constants of nature, that 
need to be determined experimentally. In the Wolfenstein 
approximation\cite{wolf} the matrix is written as:
\begin{equation}
{\begin{array}{ccc} 
1-\lambda^2/2 &   \lambda &  A\lambda^3(\rho-i\eta(1-\lambda^2/2)) \\
-\lambda &   1-\lambda^2/2-i\eta A^2\lambda^4 &  A\lambda^2(1+i\eta\lambda^2) \\
A\lambda^3(1-\rho-i\eta) &   -A\lambda^2&  1  
\end{array}}.
\end{equation}
This expression is accurate to order $\lambda^3$ in the real part and
$\lambda^5$ in the imaginary part. It is necessary to express the matrix
to this order to have a complete formulation of the physics we wish to pursue.

Non-zero $\eta$ allows for CP violation. 
CP violation thus far has only been seen in the neutral kaon 
system. By exploring CP violation in the $b$ and $c$ systems we can see
if the CKM  model works or perhaps discover new physics that 
goes  beyond the model, if it does not.

\begin{figure}[htb]
\centerline{\epsfig{figure=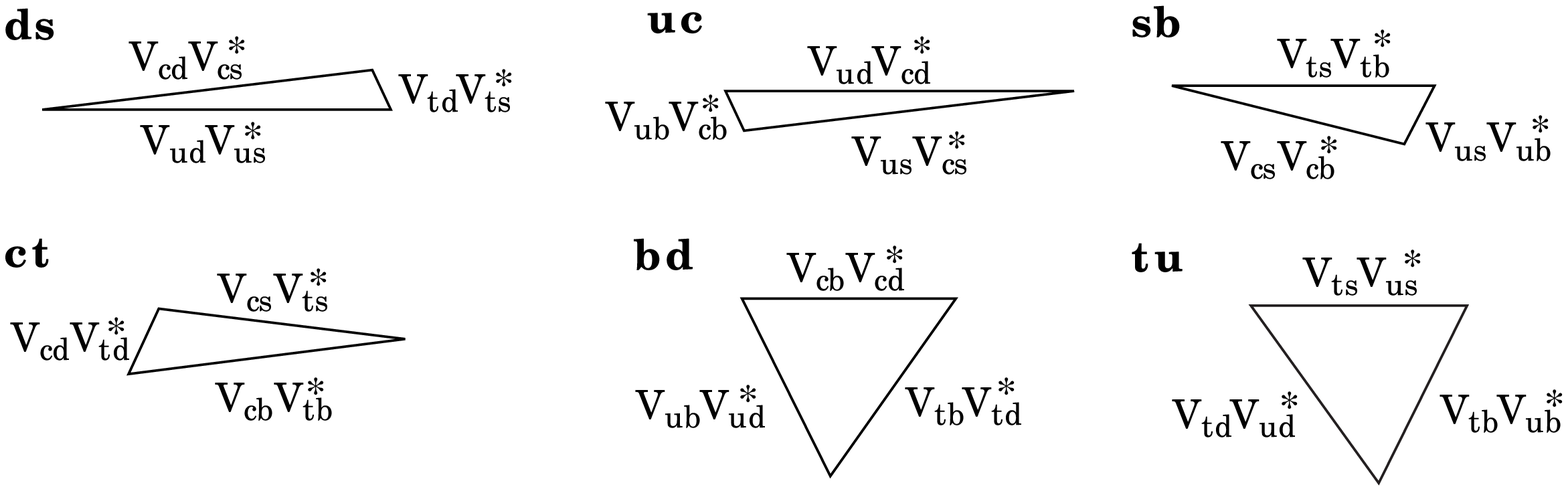,height=1.3in}}
\caption{\label{six_tri}The six CKM triangles. The bold labels, e.g. {\bf ds} 
refer to the rows or columns used in the unitarity relationship.}
\end{figure}

The unitarity of the CKM matrix allows us to construct six relationships.
These equations may be thought of triangles in the complex plane. They are 
shown in Fig.~\ref{six_tri}.

All six of these triangles can be constructed knowing four and
only four independent angles.\cite{silva_wolf}$^,$\cite{KAL}$^,$\cite{bigis}
 These 
are defined as:
\begin{eqnarray} \label{eq:chi}
\beta=arg\left(-{V_{tb}V^*_{td}\over V_{cb}V^*_{cd}}\right),&~~~~~&
\gamma=arg\left(-{{V^*_{ub}V_{ud}}\over {V^*_{cb}V_{cd}}}\right), \nonumber\\
\chi=arg\left(-{V^*_{cs}V_{cb}\over V^*_{ts}V_{tb}}\right),&~~~~~&
\chi'=arg\left(-{{V^*_{ud}V_{us}}\over {V^*_{cd}V_{cs}}}\right).\nonumber\\ 
\end{eqnarray}

Two of the phases $\beta$ and $\gamma$ are probably large while $\chi$ is
estimated to be small $\approx$0.02, but measurable, while $\chi'$ is likely
to be much smaller.

\begin{figure}[htb]
\centerline{\epsfig{figure=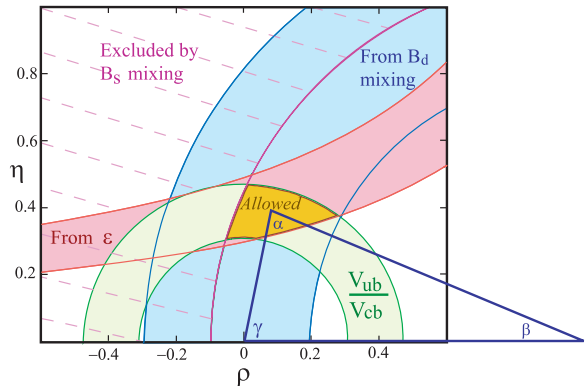,height=2in}}
\vspace{-0.2cm}
\caption{\label{ckm_tri}The CKM triangle shown in the $\rho-\eta$ plane. The
shaded regions show $\pm 1\sigma$ contours given by
$|V_{ub}/V_{cb}|$, neutral $B$ mixing, and CP violation in $K_L^o$ decay ($\epsilon$). The dashed region is excluded by $B_s$ mixing limits.
 The allowed region is defined by the overlap of
the 3 permitted areas, and is where the apex of the CKM triangle sits.}
\vspace{-4mm}
\end{figure}

In the {\bf bd} triangle, the one usually considered, the angles are all thought 
to be relatively large.
Since $V_{cd}^*=\lambda$, this triangle has sides
$1, ~~~ 
{1\over \lambda} \left|{V_{td}\over V_{ts}}\right|, ~~~
{1\over \lambda} \left|{V_{ub}\over V_{cb}}\right|.$
This CKM triangle is depicted in Fig.~\ref{ckm_tri}, with
constraints from other measurements.\cite{Stone_HF8}
Also shown are the angles
$\alpha,~\beta$, and $\gamma$. Since they form a triangle the ``real" $\alpha$,
$\beta$ and $\gamma$ must sum to 180$^{\circ}$; therefore measuring any two
of these determines the third.

It has been pointed out by Silva and Wolfenstein\cite{silva_wolf} that
measuring these angles may not be sufficient to detect
new physics. For example, suppose there is new physics that arises in 
$B^o-\overline{B}^o$ mixing. Let us assign a phase $\theta$ to this new
physics. If we then measure CP violation in $B^o\to J/\psi K_S$ and eliminate
any Penguin pollution problems in using $B^o\to\pi^+\pi^-$, then we actually
measure $2\beta' =2\beta + \theta$ and $2\alpha' = 2\alpha -\theta$. So while
there is new physics, we miss it, because
$2\beta' + 2\alpha' = 2\alpha +2\beta$ and $\alpha' + \beta' +\gamma
= 180^{\circ}$.

\subsection{Ambiguities}
In measuring CP phases there are always ambiguities. For example, any 
determination of 
$\sin(2\phi)$, has a four-fold ambiguity; $\phi$, 
$\pi/2-\phi$, $\pi+\phi$, $3\pi/2-\phi$ are all allowed solutions. Often
the point of view taken is that we know $\eta$ is a positive quantity and thus
we can eliminate two of the four possibilities. However, this is dangerous as
it could lead to our missing new physics. Evidence that $\eta$ is positive is
derived from the measurements of $\epsilon$ and $\epsilon'$ using theoretical
models. Even accepting that
$K_L$ decays give $\eta >0$, it would be foolhardy to miss new physics just because 
we now assume that $\eta$ must be positive rather than insisting on a clean 
measurement of the angles that could show a contradiction.

\subsection{Technique for Measuring $\alpha$}
\begin{figure}[htb]
\vspace{-.3cm}
\centerline{\epsfig{figure=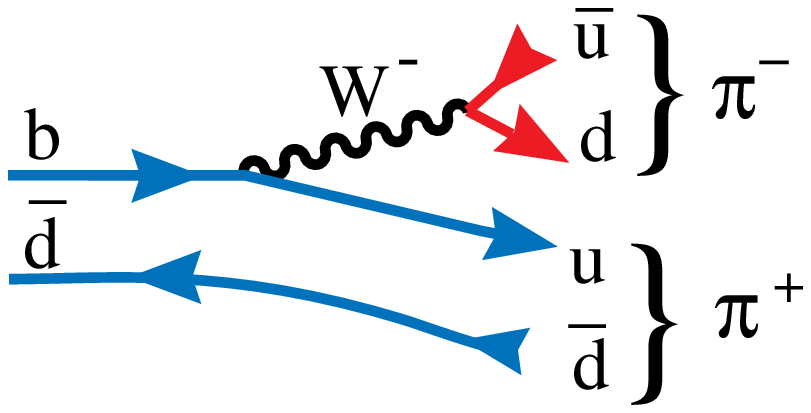,height=1.05in}\epsfig{figure=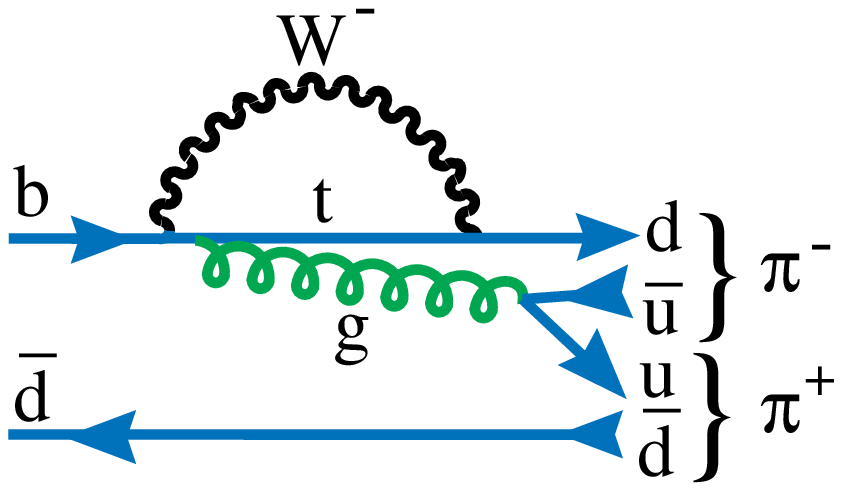,height=1.05in}}
\vspace{-0.2cm}
\caption{\label{pipi} Processes for $B^o\to\pi^+\pi^-$: Tree (left)
and Penguin (right).}
\end{figure}

It is well known that $\sin (2\beta)$ can be measured without
problems caused by Penguin processes using the reaction $B^o\to J/\psi K_S$.
The simplest reaction that can be used to measure $\sin (2\alpha)$ is
$B^o\to \pi^+\pi^-$. This reaction can proceed via both the Tree and Penguin
diagrams shown in Fig.~\ref{pipi}.

 Current CLEO results\cite{wurthwein}  are 
${\cal{B}}(B^o\to K^{\mp}\pi^{\pm})=(1.88^{+0.28}_{-0.26}\pm 0.13)\times 10^{-5}$ and 
 ${\cal{B}}(B^o\to \pi^{+}\pi^{-})=(0.47^{+0.18}_{-0.15}\pm 0.06) \times 10^{-5}$, showing a relatively large Penguin amplitude that cannot
be ignored.  The
 Penguin contribution to $\pi^+\pi^-$ is roughly half the Tree
 {\it amplitude}. Thus the effect of the Penguin must be
determined in order to extract $\alpha$. The only model independent way 
of doing this was suggested by Gronau and London, but requires the measurement
of $B^{\mp}\to\pi^{\mp}\pi^o$ and $B^o\to\pi^o\pi^o$, the latter being rather 
daunting.

There is however, a theoretically clean method to determine $\alpha$.
The interference between Tree and Penguin diagrams can be exploited by
 measuring the time dependent CP violating
 effects in the decays $B^o\to\rho\pi\to\pi^+\pi^-\pi^o$  
as shown by Snyder and Quinn.\cite{SQ}

\begin{figure}[htb]
\vspace{-0.4cm}
\centerline{\epsfig{figure=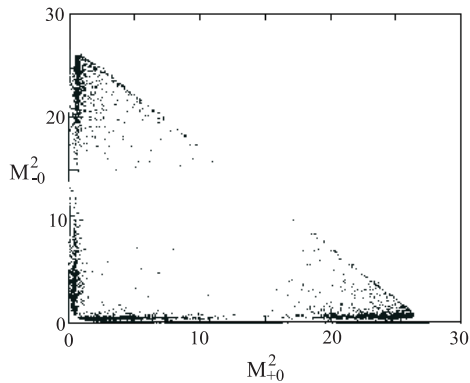,height=2.3in}}
\vspace{-.6cm}
\caption{\label{dalitz} The Dalitz plot for $B^o\to\rho\pi\to\pi^+\pi^-\pi^o$
from Snyder and Quinn.}
\end{figure}

The $\rho\pi$ final state has many advantages. First of all,
it has been seen with a relatively large rate. The 
branching ratio for the $\rho^o\pi^+$ final state as measured by CLEO\cite{CLEO_rhopi} is 
$(1.5\pm 0.5\pm 0.4)\times 10^{-5}$, and the rate for the neutral
 $B$ final state $\rho^{\pm}\pi^{\mp}$ is  
$(3.5^{+1.1}_{-1.0}\pm 0.5)\times 10^{-5}$, while the $\rho^o\pi^o$ final
state is limited at 90\% confidence level to $<5.1 \times 10^{-6}$.
These
measurements are consistent with some theoretical expectations.\cite{ali_rhopi}
Secondly, the associated vector-pseudoscalar
Penguin decay modes have conquerable or smaller branching ratios. Furthermore, since the 
$\rho$ is spin-1, the $\pi$ spin-0 and the initial $B$ also spinless, the $\rho$ 
is fully polarized in the (1,0) configuration, so it decays as $cos^2\theta$, 
where $\theta$ is the angle of one of the $\rho$ decay products with the other
$\pi$ 
in the $\rho$ rest frame. This causes the periphery of the Dalitz plot to be 
heavily populated, especially the corners. A sample Dalitz plot is shown in 
Fig.~\ref{dalitz}. This kind of distribution is good for maximizing the interferences, which 
helps minimize the error. Furthermore, little information is lost by excluding 
the Dalitz plot interior, a good way to reduce backgrounds.

To estimate the required number of events  Snyder and 
Quinn preformed an idealized analysis that showed that a background-free,
flavor-tagged sample of 1000 to
2000 events was sufficient. The 1000 event sample usually yields good results 
for $\alpha$, but sometimes does not resolve the ambiguity. With the 2000 event 
sample, however, they always succeeded. 

This technique not only finds $\sin(2\alpha)$, it also determines 
 $\cos(2\alpha)$, thereby removing two of the remaining ambiguities. The final
 ambiguity can be removed using the CP asymmetry in $B^o\to\pi^+\pi^-$ and
 a theoretical assumption.\cite{gross_quinn}

\subsection{Techniques for Measuring $\gamma$}

In fact, it may be easier to measure $\gamma$ than $\alpha$. There have been
at least four methods suggested.

(1) Time dependent flavor tagged analysis of $B_s\to D_s^{\pm}K^{\mp}$. This
is a direct model independent measurement.\cite{Aleks}

(2) Measure the rate differences between $B^-\to \overline{D}^o K^-$ and
$B^+\to {D}^o K^+$ in two different $D^o$ decay modes such as $K^-\pi^+$
and $K^+ K^-$. This method makes use of the interference between the tree
and doubly-Cabibbo suppressed decays of the $D^o$, and does not depend
on any theoretical modeling.\cite{sad}$^,$\cite{gronau}

(3) Rate measurements in two-body $B\to K \pi$ decays. A cottage industry has
developed. However, all methods are model dependent.\cite{Kpi}

(4) Use U-spin symmetry to relate $B^o\to\pi^+\pi^-$
and $B_s\to K^+ K^-$.\cite{pipiKK}

\subsection{Required Measurements Involving $\beta$}

The phase of $B^o-\overline{B^o}$ mixing will soon be measured by
$e^+e^-$ $b$-factories using the $J/\psi K_S$ final state. New physics could be
revealed by measuring other final states such as $\phi K_S$, $\eta' K_S$
or $J/\psi \pi^o$. 

It is also important to resolve the ambiguities. There are two suggestions on how
this may be accomplished. Kayser\cite{kkayser} shows that time dependent
measurements of the final state
$J/\psi K^o$, where $K^o\to \pi \ell \nu$, give a direct measurement of
$\cos(2\beta)$ and can also be used for CPT tests. Another suggestion is to use
the final state $J/\psi K^{*o}$, $K^{*o}\to K_S\pi^o$, and to compare with
$B_s\to J/\psi\phi$ to extract the sign of the strong interaction phase shift
assuming SU(3) symmetry, and thus determine $\cos(2\beta)$.\cite{isi_beta}

\subsection{A Critical Check Using $\chi$}

The angle $\chi$, defined in equation~\ref{eq:chi}, can be extracted by
measuring the time dependent CP violating asymmetry in the reaction
$B_s\to J/\psi \eta^{(}$$'^{)}$, or if one's detector is incapable of quality
photon detection, the $J/\psi\phi$ final state can be used.  However, there are
two vector particles in the final state, making this a state of mixed CP
a requiring time-dependent angular analysis to find $\chi$ that requires large statistics.

Measurements of the magnitudes of 
CKM matrix elements all come with theoretical errors. Some of these are hard 
to estimate; we now try and view realistically how to combine CP violating phase 
measurements with the magnitude measurements to best test the Standard Model.

The best measured magnitude is that of $\lambda=|V_{us}/V_{ud}|=0.2205\pm 
0.0018$. Silva and 
Wolfenstein\cite{silva_wolf}$^,$\cite{KAL}
show that the Standard Model 
can be checked in a profound manner by seeing if:
\begin{equation}
\sin\chi = \left|{V_{us}\over 
V_{ud}}\right|^2{{\sin\beta~\sin\gamma}\over{\sin(\beta+\gamma)}}~~.
\end{equation}
Here the precision of the check will be limited initially by the measurement of $\sin\chi$, not 
of $\lambda$. This check can  reveal new physics, even 
if other measurements have not shown any anomalies. There are other checks using
$\left|{V_{ub}\over V_{cb}}\right|$ or $\left|{V_{td}\over 
V_{ts}}\right|$.\cite{Stone_HF8}

\subsection{Other Critical CKM Measurements and Summary}

Magnitudes of the CKM elements are important to measure as precisely as possible.
Current measurements of $|V_{cb}|$ and $|V_{ub}|$ are discussed elsewhere.\cite{elsewhere}

It has been predicted that $\Delta\Gamma/\Gamma$ for the $B_s$ system is of
the order
of 10\%. This can be determined by measuring the lifetimes in different
final states such as $D_s^-\pi^+$ (mixed CP), $J/\psi \eta'$ (CP $-$) and
$K^+ K^-$ (CP +). A finite $\Delta\Gamma$ would allow many other interesting
measurements of CP violation.\cite{DGamma}

Table~\ref{table:reqmeas} lists the most important physics quantities and the
suggested decay modes. The necessary detector capabilities include
the ability to collect purely hadronic final states, the ability to identify
charged hadrons, the ability to detect photons with good efficiency and
resolution and  excellent time resolution required to analyze rapid $B_s$
oscillations.

\begin{table}[hbt]
\begin{center}
\caption{Required CKM Measurements for $b$'s}
\label{table:reqmeas}
\begin{tabular}{llcccc} \hline\hline
Physics & Decay Mode & Hadron & $K\pi$ & $\gamma$ & Decay \\
Quantity&            & Trigger & sep   & det & time $\sigma$ \\
\hline
$\sin(2\alpha)$ & $B^o\to\rho\pi\to\pi^+\pi^-\pi^o$ & $\surd$ & $\surd$& $\surd$ 
&\\
$\cos(2\alpha)$ & $B^o\to\rho\pi\to\pi^+\pi^-\pi^o$ & $\surd$ & $\surd$& 
$\surd$ &\\
sign$(\sin(2\alpha))$ & $B^o\to\rho\pi$ \& $B^o\to\pi^+\pi^-$ & 
$\surd$ & $\surd$ & $\surd$ & \\
$\sin(\gamma)$ & $B_s\to D_s^{\pm}K^{\mp}$ & $\surd$ & $\surd$ & & $\surd$\\
$\sin(\gamma)$ & $B^-\to \overline{D}^{0}K^{-}$ & $\surd$ & $\surd$ & & \\
$\sin(\gamma)$ & $B^o\to\pi^+\pi^-$ \& $B_s\to K^+K^-$ & $\surd$ & $\surd$& & 
$\surd$ \\
$\sin(2\chi)$ & $B_s\to J/\psi\eta',$ $J/\psi\eta$ & & &$\surd$ &$\surd$\\
$\sin(2\beta)$ & $B^o\to J/\psi K_s$ & & & & \\
$\cos(2\beta)$ &  $B^o\to J/\psi K^o$, $K^o\to \pi\ell\nu$  & & & & \\
$\cos(2\beta)$ &  $B^o\to J/\psi K^{*o}$ \& $B_s\to J/\psi\phi$  & & & 
&$\surd$ \\
$x_s$  & $B_s\to D_s^+\pi^-$ & $\surd$ & & &$\surd$\\
$\Delta\Gamma$ for $B_s$ & $B_s\to  J/\psi\eta'$, $ D_s^+\pi^-$, $K^+K^-$ &
$\surd$ & $\surd$ & $\surd$ & $\surd$ \\
\hline
\end{tabular}
\end{center}
\end{table}

\section{Searches for New Physics}

Because new physics at much larger mass scales can appear in loops, rare
process such as $b\to s\gamma$, $d\gamma$, $s\ell^+\ell^-$ and
$d\ell^+\ell^-$ have the promise to reveal new physics. Searches in both
exclusive and inclusive final states are important.

Charm decays also offer the possibility of finding new physics in the study of
either mixing or CP violation as the Standard Model prediction is small.
The current experimental measurement of mixing is $r_D < 5\times 10^{-3}$,
while the SM expectation\cite{charm_mix} is $10^{-7}-10^{-6}$. For CP violation the current
limits are about 10\%, while the expectation\cite{charm_cp} is $10^{-3}$. 

\section{The Next Generation of Experiments: LHC-b and BTeV }
\subsection{Rationale}

To over constrain the CKM matrix and look for new physics, all the quantities
listed in Table~\ref{table:reqmeas} must be measured. This requires large
samples of $b$-flavored hadrons, and detectors capable of tolerating large
interaction rates and having excellent lifetime resolution, particle
identification and $\gamma/\pi^o$ detection capabilities. 

Fortunately, large samples of $b$ quarks are available. With the Fermilab Main
Injector, the Tevatron collider will produce $\approx 4\times 10^{11}$  $b$ 
hadrons/$10^7$ s at a luminosity of $2\times 10^{32}$cm$^{-2}$s$^{-1}$. Rates are $\approx 5$ times larger at  the LHC. These compare very favorably to
e$^+$e$^-$ machines operating at the $\Upsilon$(4S). At a luminosity of 
$3\times 10^{33}$ they produce $6\times 10^7$ B's/$10^7$ s. Furthermore
$B_s$, $\Lambda_b$ and other $b$-flavored hadrons are accessible for study at hadron colliders. Also important are the large
charm rates, $\sim$10 times larger than the $b$ rate.

\subsection{$b$ Production Characteristics}
\begin{figure}[htb]
\vspace{-1.cm}
\centerline{\epsfig{figure=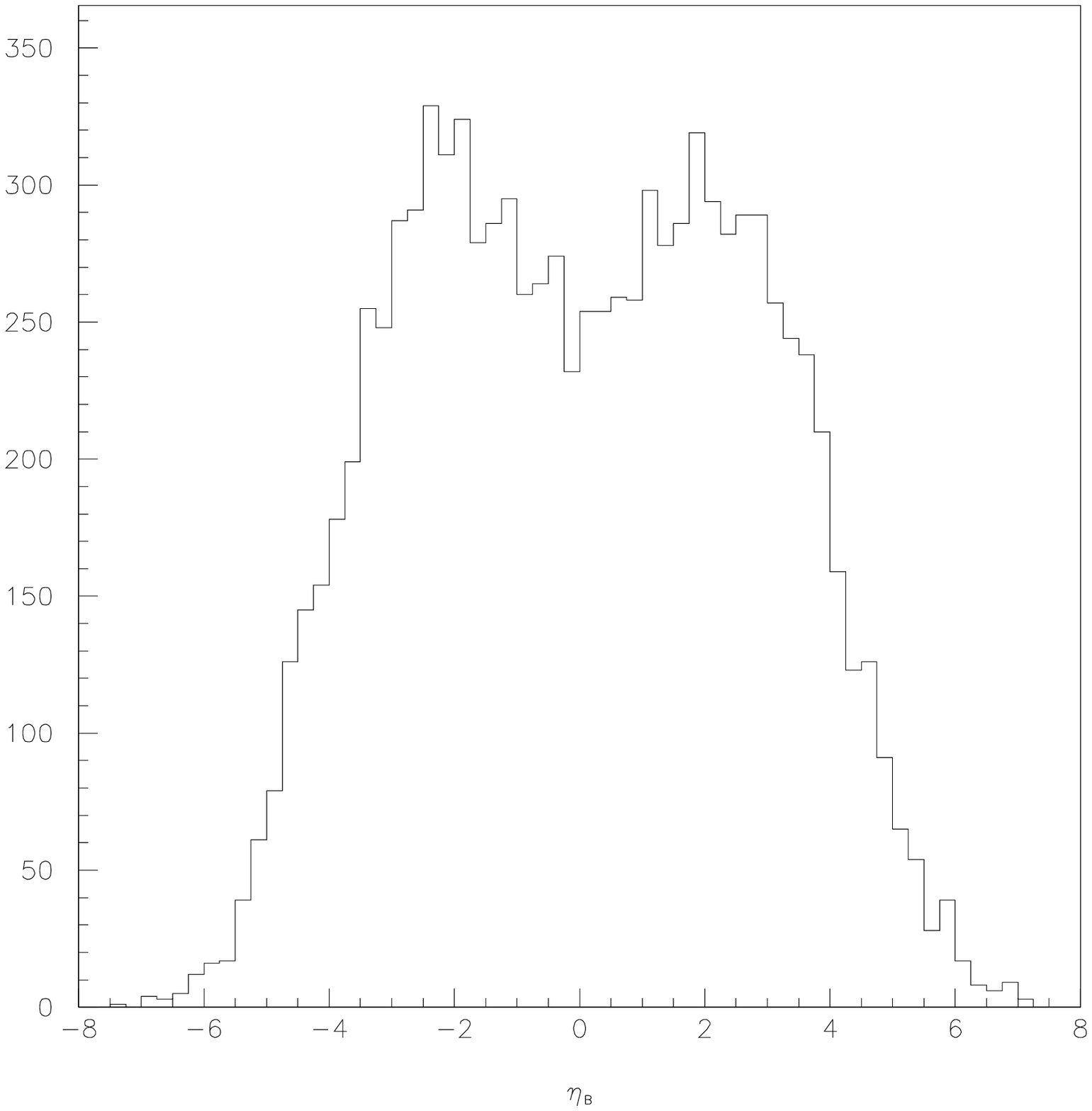,width=1.75in}
\epsfig{figure=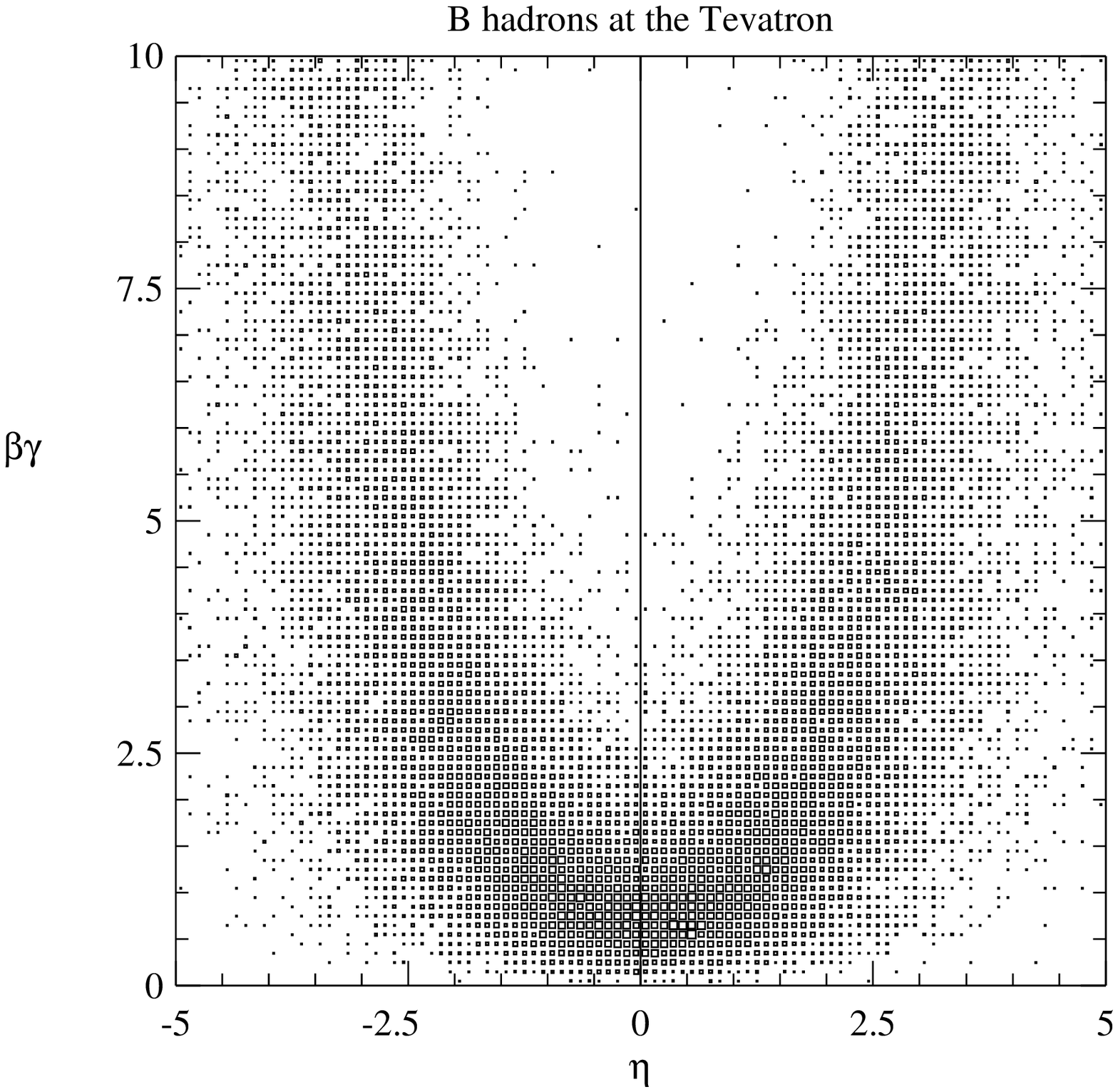,width=1.82in}}
\vspace{-.2cm}
\caption{\label{n_vs_eta}  The $B$ yield versus $\eta$ (left). 
$\beta\gamma$ of the  $B$  versus $\eta$ (right). Both plots are for the
Tevatron.}
\end{figure}

Let us explore the reasons for the choice of the forward direction. It is often
customary to characterize heavy quark production in hadron collisions with the
two variables $p_t$ and $\eta$, where  $\eta =
-ln\left(\tan\left({\theta/2}\right)\right),$ and $\theta$ is the angle of the
particle with respect to the beam direction. According to QCD based
calculations of $b$ quark production, the $B$'s are produced ``uniformly" in
$\eta$ and have a truncated transverse momentum, $p_t$, spectrum characterized
by a mean value approximately equal to the $B$ mass.\cite{artuso} The
distribution in $\eta$ is shown in Fig.~\ref{n_vs_eta}. Note that at larger
values of $|\eta|$, the $B$ boost, $\beta\gamma$, increases rapidly.

\begin{figure}[hbt]
\centerline{\epsfig{figure=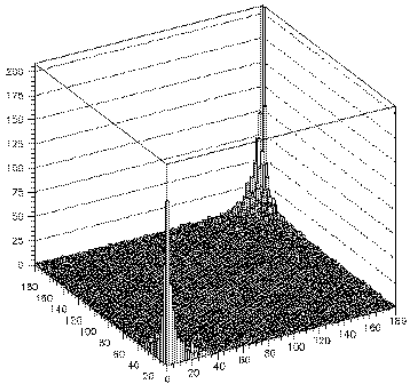,height=2.in}}
\caption{\label{bbar}The production angle (in degrees) for the hadron
containing a $b$ quark plotted versus the production angle for a hadron
containing $\bar{b}$ quark. (For the Tevatron.) $\eta$ of zero corresponds
to 90$^o$ here.}
\end{figure}

The ``flat" $\eta$ distribution hides an important correlation of
$b\bar{b}$ production at hadronic colliders. In Fig.~\ref{bbar} the production
angles of the hadron containing the $b$ quark is plotted versus the production
angle of the hadron containing the $\bar{b}$ quark according to the Pythia
generator. Many important measurements require the reconstruction of a $b$ decay and 
the determination of the flavor of the other $\bar{b}$, thus requiring both $b$'s to be 
observed in the detector. There is a very strong
correlation in the forward (and backward) direction: when the $B$ is forward
the $\overline{B}$ is also forward. This correlation is not present in the
central region (near zero degrees). By instrumenting a relative small region of
angular phase space, a large number of $b\bar{b}$ pairs can be detected. 
Furthermore the $B$'s populating the forward and backward regions have large
values of $\beta\gamma$. 

Both experiments use forward spectrometers which utilizes the boost of the
$B$'s at large rapidities. This is of crucial importance because the main way
to distinguish $b$ decays is by the separation of decay vertices from the main
interaction vertex.

\subsection{General Design Considerations}

Two dedicated experiments are
contemplated, LHC-b which has been approved and BTeV, an approved R\&D program
at Fermilab that was requested to prepare a proposal for submission in summer 
of 2000.\cite{harnew} They both look for $b$ decays in the ``forward" 
direction close to the beam line to exploit the large $b$ momenta and the 
correlated $b\overline{b}$ production. As a result, they both have long and 
narrow interaction regions. The C0 interaction region at Fermilab was constructed to allow the BTeV experiment to fit.

There are problems that heavy quark experiments at hadron colliders must overcome. First of all, the hugh $b$ rate is accompanied by an even larger rate of uninteresting interactions. At the Tevatron the $b$-fraction is only 1/500, and is only 5 times larger at the LHC. In searching for rare processes, at the level of parts per million, the background from $b$ events is dominant. (Of course all $b$ experiments have this problem.) The large data rate of $b$'s must be handled. For example, BTeV, has 1 kHz into the detector, and these events must be selected and written out. The electromagnetic calorimeter must robust enough to deal with the particles from the underlying event and still have useful efficiency. Furthermore, radiation damage can destroy detector elements.

The strategy BTeV employes to overcome these obstacles includes triggering on 
events with detached vertices in the first trigger level. Both BTeV and LHC-b 
use the excellent detached vertex resolution,  $\sim40$ fs, to reject 
backgrounds in 
their analyses. Both experiments incorporate deadtimeless trigger and data 
acquisition systems, and have excellent Cherenkov Ring Imaging detectors to 
select charged hadrons. Furthermore BTeV has an excellent Electromagnetic 
calorimeter made from PbWO$_4$ crystals, based on the design of CMS. Both 
experiments can measure $B_s$ mixing up to $x_s$ values of 60-80.

LHC-b employs a somewhat different trigger strategy. They have a first level 
trigger (which they somewhat annoyingly call level-0) that selects events on 
the basis of having hadronic tracks or leptons with transverse momenta in the 
range of a few GeV/c. They then ask for detached verticies in the next trigger 
level. 

\subsection{Short Detector Descriptions}
\begin{figure}[hbt]
\centerline{\epsfig{figure=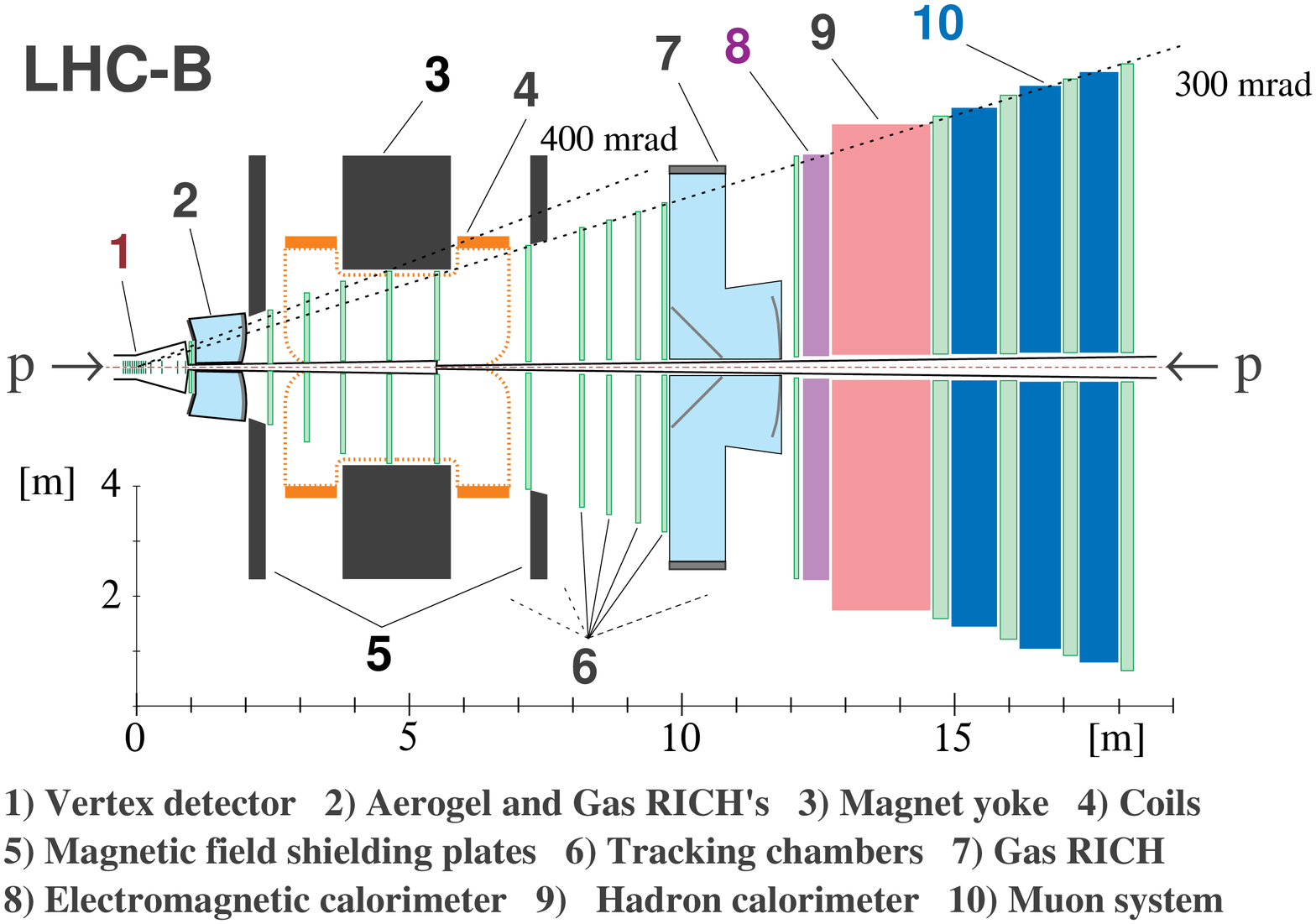,height=2.8in}}
\vspace{-.9cm}
\caption{\label{LHC_B}Schematic of the LHC-b detector.}
\end{figure}

The LHC-b detector is shown in Fig.~\ref{LHC_B}. The vertex detector is made 
from silicon strips. It is located in a magnetic field free region, and placed as close as 10 mm from the beam line. Two Ring Imaging Cherenkov detectors are used with different gases to cover the full momentum range for $K/\pi$ separation. In addition, to separate kaons from protons, an aerogel radiator will likely be attached to the entrance window on the first RICH. There is a ``Shaslik" style electromagnetic calorimeter made from scintillating fibers embedded in lead and a hadron calorimeter whose primary use is to veto crossing with more than one interactions, since these can confuse the trigger.

\begin{figure}[hbt]
\centerline{\epsfig{figure=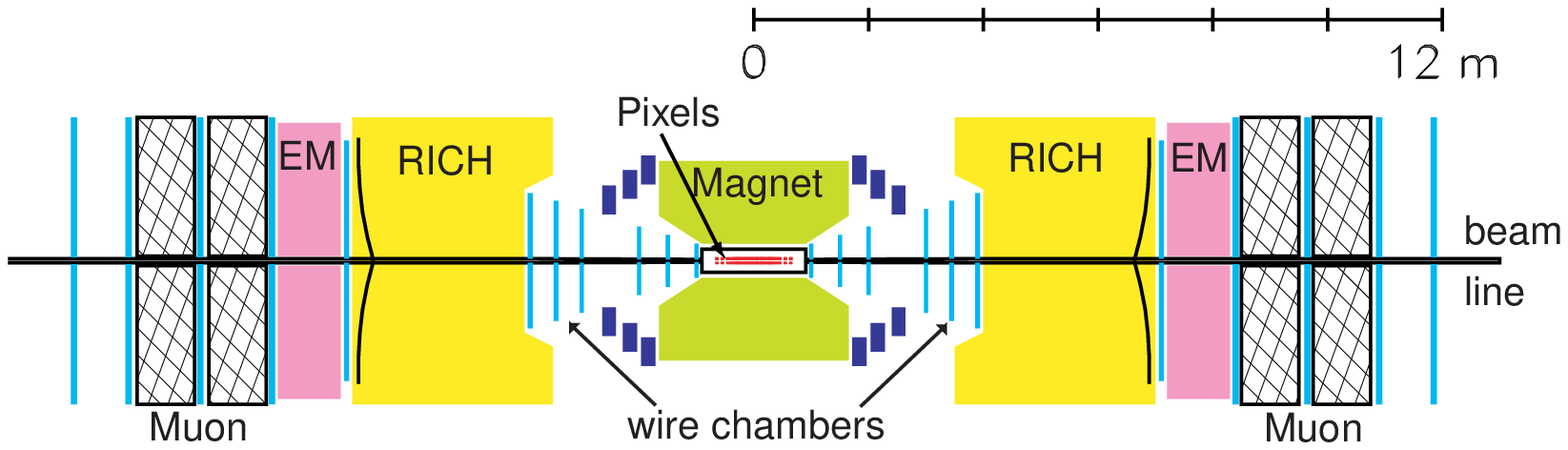,height=1.4in}}
\caption{\label{btev_det}Schematic of the BTeV detector.}
\end{figure}

The BTeV Detector is shown in Fig.~\ref{btev_det}. It has two ``forward" arms. This is done to increase the accepted $b$ rate. BTeV must make up a factor of $\approx$5 to compete with LHC-b. BTeV has the vertex detector in the magnetic field. It is a pixel detector and is placed 6 mm from the beam line. BTeV believes that the detached vertex triggering is enhanced by eliminating low momentum, large multiple scattering tracks, from consideration. The single RICH, also with aerogel, is followed by a PbWO$_4$ electromagnetic calorimeter based on the design of CMS for radiation hard crystals. The crystals are approximately 26 mm x 26 mm x 220 mm. This gives good segmentation. Note that the lower momenta and multiplicities allow each arm of BTeV to be about half the length of LHC-b.

\subsection{Some Sensitivity Projections}
\begin{figure}[hbt]
\vspace{-.4cm}
\centerline{\epsfig{figure=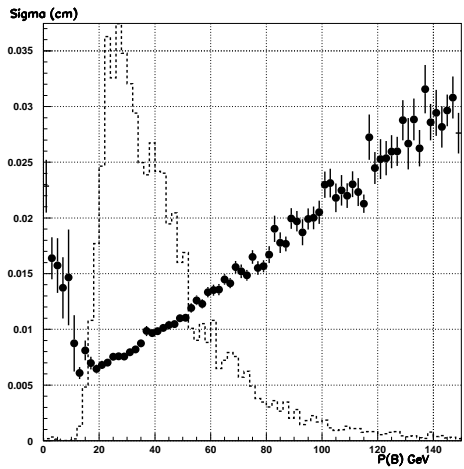,height=3.in}}
\caption{\label{pipi_decay}The momentum distribution (dashed) of $B$'s accepted by BTeV
in the $\pi^+\pi^-$ decay mode along with the error on the vertex position
measurment (solid points).}
\end{figure}

\begin{figure}[hbt]
\centerline{\epsfig{figure=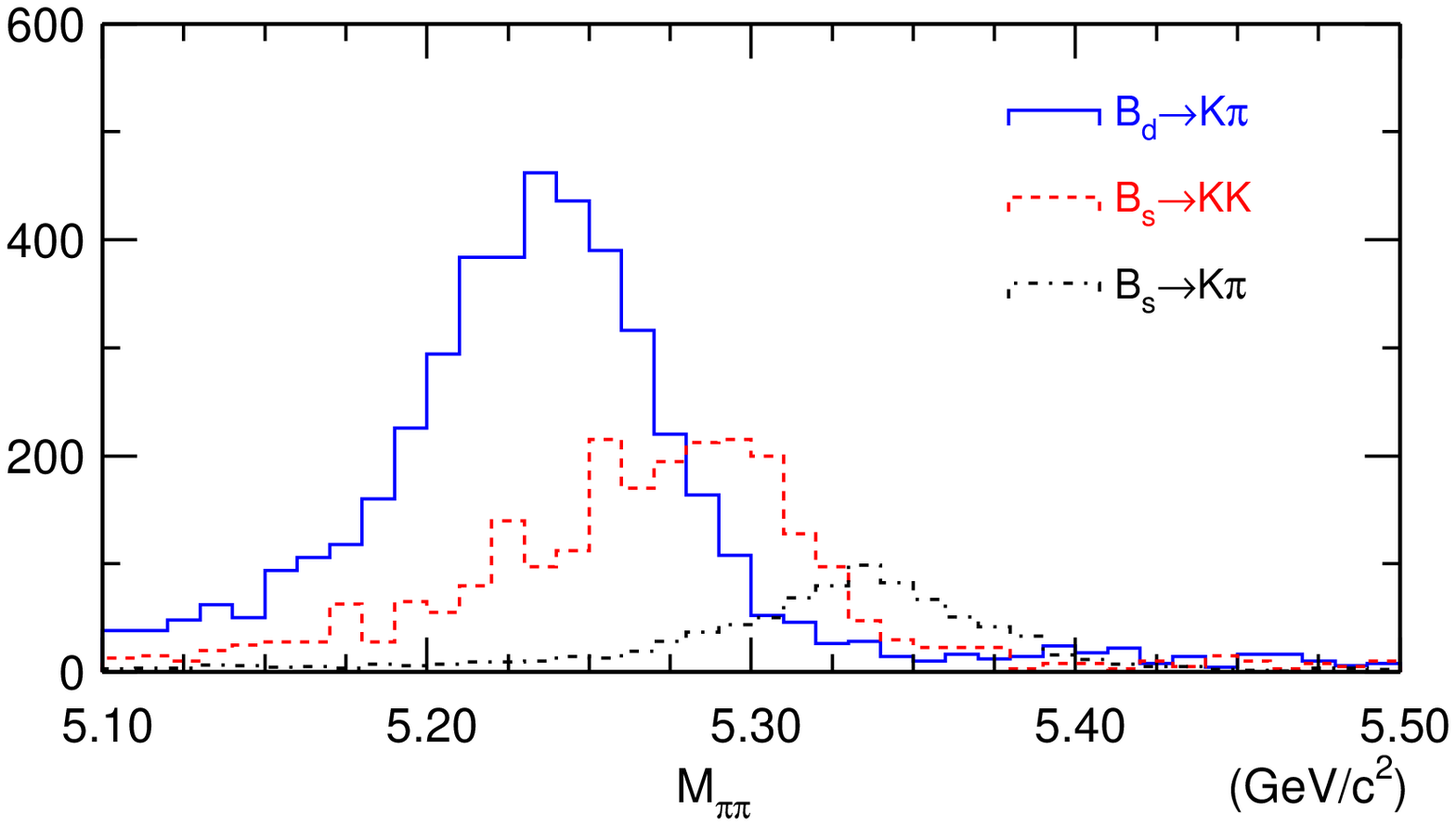,height=1.4in}
\epsfig{figure=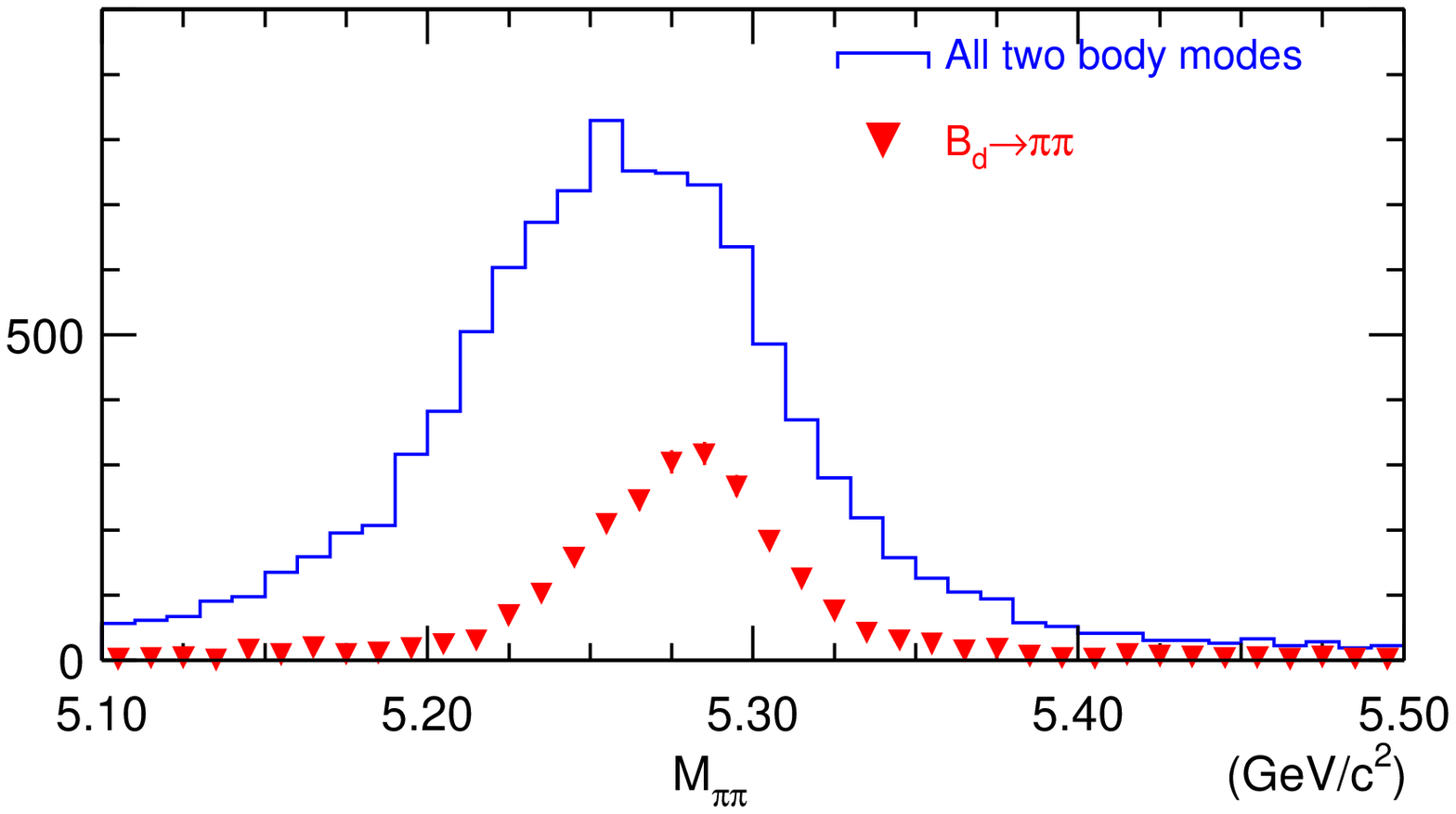,height=1.35in} }
\caption{ \label{b_pipi_pide}Two-body mass plots without particle 
identification left) 
$B_d \rightarrow K^+\pi^- $,
$B_s \rightarrow \pi^+K^- $, $B_s \rightarrow K^+K^- $, 
right) $B_d \rightarrow
\pi^+\pi^-$ and a sum of all two body decay modes.  All particles are
assumed to be pions.  }
\end{figure}

Space precludes a through discussion of sensitivities. I will discuss one 
illustrative example, the CP asymmetry in $B^o\to\pi^+\pi^-$. I use a BTeV 
simulation. The $B^o$ momentum distribution for events in the detector is 
shown in Fig.~\ref{pipi_decay}. Also shown is the error in decay distance. 
The peak of the $B^o$ momentum distribution is about 30 GeV/c. Since the 
average decay length goes as 480 $\mu$m$\times p_B/m_B$, 30 GeV/c $B$'s go 
about 3 mm.  Below about 20 GeV/c the error on measuring the 
decay distance grows, while above the decay distance error grows linearly 
with $B$ momentum. The error growth at low momentum is due to multiple 
scattering. Since the key variable is decay distance divided by the error on 
decay distance, $L/\sigma$, it is best to have $B$'s above 20 GeV/c. 
$L/\sigma$ is key in triggering as well as rejecting background. However, 
another significant source of background could be the other two-body decays 
into two pseudoscalars. These include $B^o\to K^{\pm}\pi^{\mp}$ and $K^+K^-$, 
and $B_s\to K^+K^-$ and $ K^{\pm}\pi^{\mp}$. Fig.~\ref{b_pipi_pide} shows the reconstructed 
mass spectra using the current CLEO measured branching ratios\cite{wurthwein} 
for $K^-\pi^+$ and $\pi^+\pi^-$, and assuming the corresponding modes in 
$B_s$ decay have the same branching fractions.
The $\pi^+\pi^-$ signal is completely swamped. However, the RICH particle 
identification should remove almost all the background. The relevant 
sensitivity calculations are shown for LHC-b and BTeV in Table~\ref{table:Btopipi}
\begin{table}[hbt]
\centering
\caption{Sensitivity for the CP Asymmetry in $B^o\to\pi^+\pi^-$ }
\label{table:Btopipi}
\vspace*{2mm}
\begin{tabular}{|l|c|c|} \hline
Quantity & BTeV & LHC-b\\\hline
$b$ cross-section & 100 $\mu$b& 500 $\mu$b\\
Luminosity & $2\times 10^{32}$cm$^{-2}$s$^{-1}$ & $2\times 10^{32}$cm$^{-2}$s$^{-1}$\\
\# $B^o$/Year ($10^7$ s) & $1.4\times 10^{11}$ & $7.0\times 10^{11}$ \\
${\cal{B}}(B^o\to \pi^+\pi^-)$ & $0.47\times 10^{-6}$ &  $0.47\times 10^{-6}$ \\
Reconstruction efficiency & 0.06 & 0.03 \\
Triggering efficiency  & 0.50 & 0.17 \\
\# of $(\pi^+\pi^-)$ &20,000  & 18,700 \\
$\epsilon D^2$ for flavor tags &0.1 &0.1 \\
Signal/Background &0.4 &0.67\\
Error in Asymmetry  & $\pm$0.042 &$\pm$0.037 \\
\hline
\end{tabular}
\end{table}

Comparing with $e^+e^-$ machines at a luminosity of $3\times 10^{33}$, the 
sensitivity is vastly greater; there are only 13 flavor tagged $\pi^+\pi^-$ 
events in $e^+e^-$.

\section{Conclusions}
Both LHC-B and BTeV will be able to make what we now consider to be the
quintessential measurements in $b$ and charm 
physics.
The sensitivities on the angle $\gamma$ are estimated to be better than
10$^{\circ}$. It is 
likely that at least two of the ambiguities in $\sin(2\beta)$ will be 
resolved. To measure $\alpha$ the reaction $B^o\to\rho\pi$ looks most
promising. Here preliminary estimates give BTeV a factor of 50 larger
efficiency. However, neither experiment has estimated the backgrounds.
Both experiments aim toward measuring $\chi$.
BTeV, with its excellent electromagnetic calorimeter will use primarily the 
$B_s\to J/\psi \eta'$ mode, while both experiments will also use the 
$J/\psi\phi$ final state. Besides the precise measurement of CKM angles and 
resolving ambiguities, the search for new physics will be of paramount 
importance.

\section{Acknowledgments}
Support for this work was provided by the National Science Foundation.
I thank George Hou and Hai-Yang Cheng for organizing an excellent and
fruitful meeting. I thank Marina Artuso, Boris Kayser and Tomasz Skwarnicki
for many useful discussions.


\begin{thebibliography}{99}
\bibitem{ckm} N. Cabibbo, {\it Phys. Rev. Lett.} {\bf 10}, 531 (1963);
M. Kobayashi and K.~Maskawa, {\it Prog. Theor. Phys.} {\bf 49}, 652 (1973).

\bibitem{wolf}
L. Wolfenstein, {\it Phys. Rev. Lett.} {\bf 51}, 1945 (1983).

\bibitem{silva_wolf}
J. P. Silva, L. Wolfenstein, \Journal{\PRD}{55}{5331}{1997} [hep-ph/9610208].

\bibitem{KAL}
R. Aleksan, B. Kayser and D. London, \Journal{\PRL}{73}{18}{1994} [hep-ph/9403341].

\bibitem{bigis}
I. I. Bigi and A. I. Sanda, ``On the Other Five KM Triangles,'' [hep-ph/9909479].

\bibitem{Stone_HF8}
S. Stone, ``Future of Heavy Flavour Physics: Experimental Perspective," presented at Heavy Flavours 8, Southapton, UK, 1999, [hep/ph xxxx].


\bibitem {Soniepsp} 
T. Blum \etal (RIKEN-BNL-Columbia), hep-lat/9908025.

\bibitem{wurthwein}
Y. Kwon \etal ~(CLEO), ``Study of Charmless Hadronic $B$ Decays into the
Final States $K\pi$, $\pi\pi$ and $KK$, with the First Observations of
$B^o\to\pi^+\pi^-$and $B^o\to K^o\pi^o$,'' Conf 99-14, [hep-ex/9908039].

\bibitem{SQ}
A. E. Snyder and H. R. Quinn, Phys. Rev. {\bf D 48}, 2139 (1993).

\bibitem{CLEO_rhopi}
Y. Gao and F. W\"urthwein, CLEO preprint [hep-ex/9904008].

\bibitem{ali_rhopi}
 A. Ali, G. Kramer, and C.D. Lu,
    {\it Phys. Rev.} {\bf D 59}, 014005 (1999) [hep-ph/9805403].  
  
\bibitem{gross_quinn}
Y. Grossman and H. R. Quinn, \Journal{\PRD}{56}{7259}{1997}  [hep-ph/9705356].

\bibitem{Aleks}
D. Du, I. Dunietz and Dan-di Wu, {\it Phys. Rev} {\bf D 34}, 3414 (1986);
R. Aleksan, I Dunietz, and B. Kayser, {\it Z. Phys.} {\bf C 54}, 653 (1992);
R. Aleksan, A. Le Yaouanc, L. Oliver, O. P\`ene
    and J.-C. Raynal, Z. Phys. {\bf C67} (1995) 251 [hep-ph/9407406]. 

\bibitem{sad}
D. Atwood, I. Dunietz and A. Soni, \Journal{\PRL}{78}{3257}{1997}.

\bibitem{gronau}
M. Gronau and D. Wyler, \Journal{\PLB}{265}{172}{1991}.

\bibitem{Kpi}
R. Fleischer, and T. Mannel, {\it Phys. Rev.} {\bf D 57}, 2752 (1998) 
[hep-ph/9704423]; 
M. Neubert and J. L. Rosner, {\it  Phys. Rev. Lett.} {\bf  81}, 5076 (1998) 
[hep-ph/9809311];
M. Gronau, and J. L. Rosner, {\it Phys. Rev.} {\bf D 57},6843 (1998) 
[hep-ph/9711246];
M. Gronau, and D. Pirjol,  [hep-ph/9902482];
M. Gronau and J. L. Rosner, \Journal{\PRD}{59}{113002}{1999}  [hep-ph/9809384];
J.-M. Gerard and J. Weyers, {\it Eur. Phys. J} {\bf C7}, 1 (1999) 
[hep-ph/9711469].

\bibitem{pipiKK}
R. Fleischer, \Journal{\PLB}{459}{306}{1999} [hep-ph/9903456].

\bibitem{kkayser} 
B. Kayser, ``Cascade Mixing and the CP-Violating Angle Beta,''
[hep-ph/9709382]. Previous work in this area was done by Y. Azimov,
{\it Phys. Rev.} {\bf D 42}, 3705 (1990).

\bibitem{isi_beta}
A. Dighe, I. Dunietz and R. Fleischer, \Journal{\PLB}{433}{147}{1998} [hep-ph/9804254].

\bibitem{elsewhere}
Marina's EPS

\bibitem{DGamma}
For a recent calculation of $\Delta\Gamma_s$, see 
M. Beneke, G. Buchalla, C. Greub, A. Lenz and U. Nierste, 
\Journal{\PLB}{459}{631}{1999} [hep-ph/9808385]. See also
I. Dunietz, {\it Phys.\ Rev.} {\bf D 52}, 3048 (1995).

\bibitem{charm_mix}
H. Georgi, \Journal{\PLB}{297}{353}{1992}; T. Ohl \etal , \Journal{\NPB}{403}{603}{1993}.

\bibitem{charm_cp}F. Buccella \etal , \Journal{\PLB}{302}{319}{1993};
F. Buccella \etal , \Journal{\PLB}{379}{249}{1996}. 

\bibitem{artuso} M. Artuso,``Experimental Facilities for b-Quark Physics,'' in {\em $B$ Decays} revised
2nd Edition, Ed. S. Stone, World Scientific, Sinagapore (1994).
\bibitem{harnew}
Harnew has recently discussed the prospects for these experiments and
CMS and ATLAS as well. See
N. Harnew, ``Prospects for LHC-b, BTeV, ATLAS, CMS," in proceedings of
HF8.

\end{thebibliography}
\end{document}